\documentstyle[12pt]{article}
\input epsf

\begin{document}
\null\vspace{-62pt}
\begin{flushright}
hepth/9510207\\
UCSBTH-95-29\\
\end{flushright}

\centerline{\large \bf{Black Hole Condensation and Duality in String Theory
\footnote{Talk
presented at the conference,
{\sl $S$-Duality and Mirror Symmetry in String Theory}, Trieste, June,
1996.}}}
\vskip .26 in
\centerline{Andrew Strominger\footnote{e-mail:
andy@denali.physics.ucsb.edu}}

\centerline{\sl Department of Physics,}
\centerline{\sl  University of California}
\centerline{\sl Santa
Barbara, CA 93106-9530}

\vskip .50 in
The classical, four-dimensional theories derived by Calabi-Yau
compactification of string theory bear a striking resemblance to the
real world \cite{chsw}. However, these classical theories are beset by
several serious difficulties:
\begin{enumerate}
\item There are too many of them.  This is aesthetically displeasing
because a unified theory should be unique. It also entails a loss of
predictive power.
\item The theory breaks down and develops naked singularities at certain
``conifold'' points in the moduli space of the massless
four-dimensional scalar fields \cite{CDGP91}.
\end{enumerate}

In this talk we shall argue, in the context of type II string theories, that
these problems are in part resolved by nonperturbative quantum
effects. Thus --- unlike e.g.~nonabelian gauge theories --- string
theory needs quantum mechanics for consistency. This suggests that the
fundamental formulation of quantum string theory may not take the usual
form which begins with a
classical theory followed by quantization. Rather string theory
may be intrinsically
quantum in nature and not have a consistent classical limit.

The structure of conifold singularities is an old and beautiful subject
in algebraic geometry. The mathematical description will not be repeated
here. Relevant aspects and references can be found
in \cite{mbh}.  The basic picture is as follows.
The space of Calabi-Yau  string vacua  is the moduli space of Ricci-flat
metrics on the Calabi-Yau. For each coordinate $Z^i$ on the moduli space
there is a massless 4D scalar $Z^i(x)$ which describes how the
size and shape of the Calabi-Yau vary in spacetime. These moduli fields
are governed by the 4D effective action
\begin{equation}
{\cal L}_{\rm eff}= \int d^4
x\ \sqrt{-g}\ G_{ij}(Z) \nabla_\mu Z^i \nabla_\nu Z^j g^{\mu\nu}\ .
\label{msm}
\end{equation}
where $g$ is the metric on spacetime and $G$ is the metric on the moduli
space.

The moduli space metric $G$ is classically determined from Calabi-Yau
data \cite{yssc}.  In the (type II) context which we consider, there are
no quantum corrections to $G$ due to $N=2$ supersymmetric
nonrenormalization theorems.
\vskip .13 in
\centerline{\epsfysize=3.00in
\epsfbox{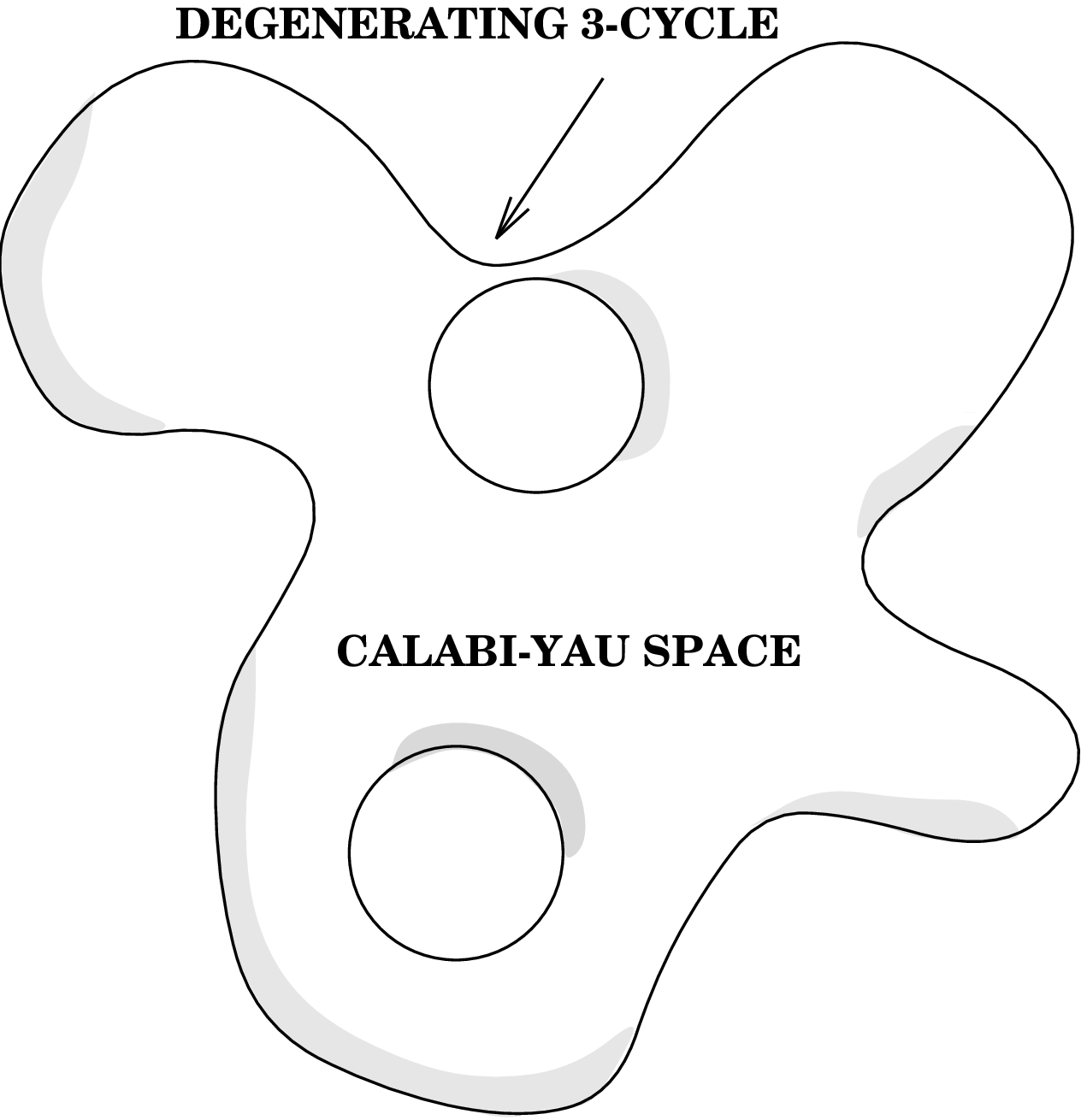}}

\begin{quote}\setlength{\baselineskip}{.1in}{\bf Figure 1}: \sl Near a
conifold, a minimal 3-cycle degenerates to
 zero volume and the Calabi-Yau space
develops a singular node. (The handles are meant to indicate the
complex topology involved: real Calabi-Yau spaces have $\pi_1=0$.)
\end{quote}

The $Z^i$'s measure the size of topologically non-trivial,
minimal-volume cycles (i.e. submanifolds) embedded in
the Calabi-Yau \cite{bbs}.
To be definite, let us consider minimal 3-cycles. At ``conifold'' points
in the moduli space, the minimal volume of
a topologically non-trivial 3-cycle can actually shrink to zero. We
can choose local coordinates so that the conifold singularity is at
$Z^1=0$. At
$Z^1=0$ the Calabi-Yau develops a singular node and is no longer a
smooth manifold as depicted in figure 1. Conifold singularities
generically occur at finite distances in the moduli space.

It is perhaps not surprising that the moduli space metric $G$ in (\ref{msm})
itself turns out to be logarithmically singular at $Z^1=0$ \cite{CDGP91}
\begin{equation}
G \sim \ell n|Z^1|\ .
\label{gsng}
\end{equation}
This is a real curvature singularity and cannot be eliminated by a
coordinate transformation of the $Z^i$'s.
 Thus
classical string theory breaks down whenever a moduli field happens to
run into a conifold singularity. It can be seen that these singularities
are real codimension two so it is hard to avoid such collisions.

A curvature singularity is not the only suspicious behavior of type II
string theory near a conifold. These theories have extremal, charged
black holes whose mass can be exactly determined using $N=2$
supersymmetry.  These masses are proportional to
\begin{equation}
M_{\rm BH} = |Z^1|\ .
\label{mbhf}
\end{equation}
Hence the black hole becomes massless at the conifold singularity
$Z^1=0$.\footnote{\setlength{\baselineskip}{.1in}
Far from the conifold the black holes are
well-described by semiclassical solutions with horizons. However, in a
neighborhood of the conifold, its Compton wavelength exceeds its
Schwarzchild radius and the semiclassical description breaks down.}

The black hole degenerates to zero mass for a simple reason. It began
life in ten dimensions as a black 3-brane \cite{hst}. This is an extended
black hole whose horizon is topologically $R^3\times S^5$, and with a
constant mass per unit three volume. In a Calabi-Yau compactification
these 3-branes can wrap around a non-trivial 3-cycle. To a low-energy 4D
observer, such a configuration will appear to be an ordinary extremal
black hole with mass proportional to the volume of the 3-cycle. When
this volume degenerates at a conifold the 4D mass will degenerate
along with it.

To summarize the picture so far, the conifold is characterized by
\begin{eqnarray}
Z^1 &\to&0\ ,\nonumber\\
M_{\rm BH} &\to&0\ ,\nonumber\\
{\cal L}_{\rm eff}&\to&\infty\ .
\end{eqnarray}
In fact this situation is not as disturbing or unusual as it seems.  It
is well known that
massless particles produce singularities in
low-energy effective actions due to infrared divergent loop
integrations. A non-singular description of the physics
can be found in a Wilsonian effective action ${\cal L}^w_{\rm eff}$.
This is obtained (in
principle) by starting from the exact microscopic theory and integrating
out fluctuations of all fields --- massive and massless --- down to some
Wilsonian cutoff $M_c$, well below the Planck or string scales,
 as depicted in figure 2.  This action differs from the
 1PI (one-particle-irreducible) effective action ${\cal L}_{\rm eff}$
usually discussed in string
theory in which fluctuations of all wavelengths are integrated out.
Divergences in the 1PI effective action arise in integrating out fluctuations
of massless fields from $M_c$ down to zero energy. Computations of a
scattering process with external momenta of order $p$ using
${\cal L}^w_{\rm eff}$ involves a quantum loop expansion with loop momenta
cutoff at $M_c$. Infrared divergences will then typically be controlled by
the external momenta $p$, and the computation will yield a finite
answer.
\vskip .13 in
\centerline{\epsfysize=3.00in \epsfbox{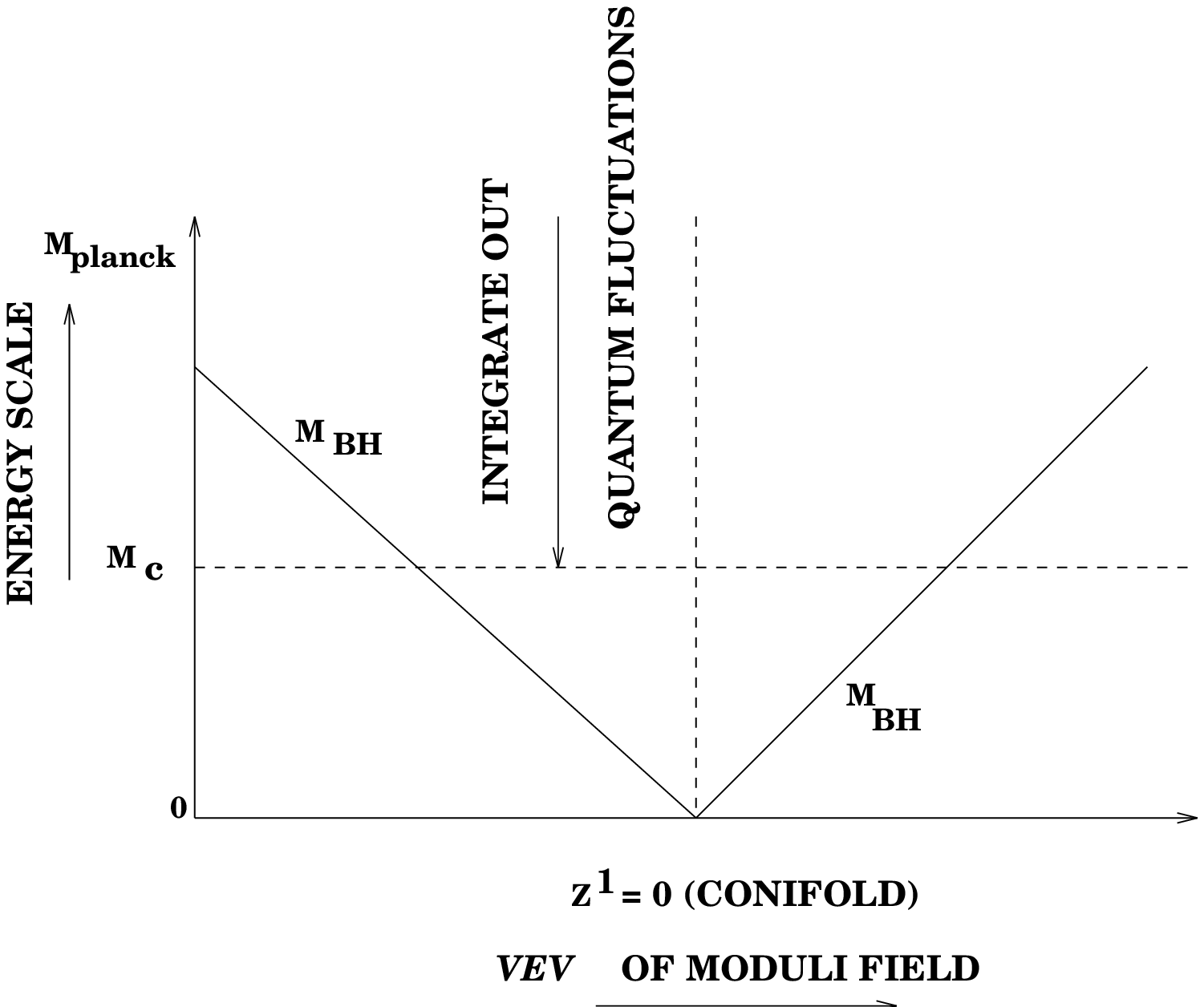}}

\begin{quote}\setlength{\baselineskip}{.1in}{\bf Figure 2}: \sl The
smooth Wilsonian effective action is defined by integrating out quantum
fluctuations of all fields down to the cutoff $M_c$. No matter how low
$M_c$ is, there is always a region in the moduli space surrounding the
conifold in which black holes are lighter than $M_c$ and
 must be included in the Wilsonian action.

\end{quote}

In the case of conifold singularities, the divergences in ${\cal L}_{\rm
eff}$ have precisely the right coefficients to have been produced by
integrating out a black hole \cite{mbh}. This has remarkably been
confirmed even for subleading terms in ${\cal L}_{\rm eff}$ \cite{vfa}.
We conclude that the underlying Wilsonian effective action has couplings
which are nonsingular as $Z^1\to0$. Finite-momentum processes
can be computed at the conifold utilizing ${\cal L}^w_{\rm eff}$. Hence
we see that classical inconsistencies of string theory are
cured by quantum loops of black holes. It is fascinating that the
demand for a
consistent theory forced us to include these black holes with virtual
fluctuations on the same footing as elementary strings.
\vskip .13 in
\centerline{\epsfysize=3.00in \epsfbox{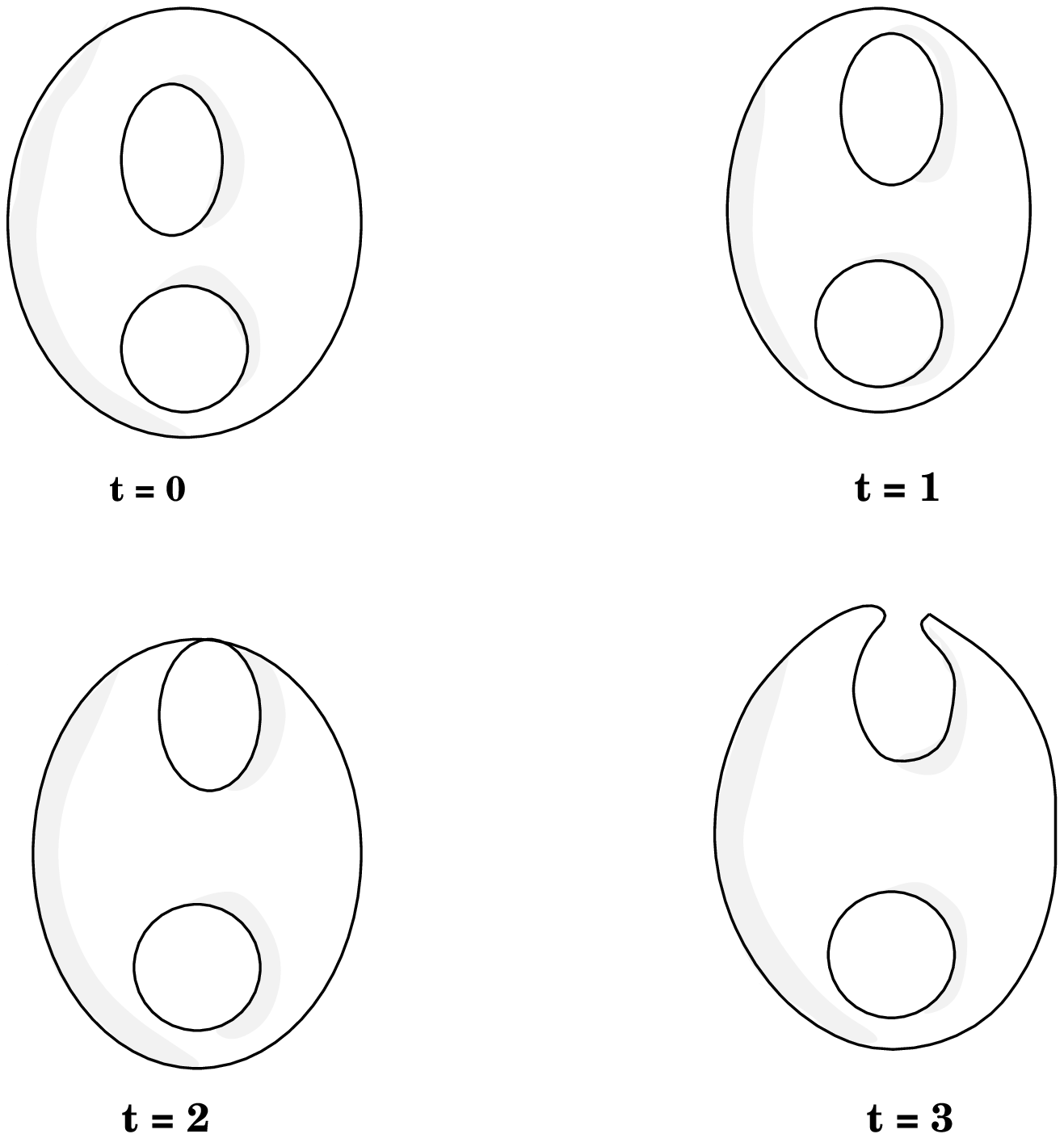}}

\begin{quote}\setlength{\baselineskip}{.1in}{\bf Figure 3}: \sl The
shape
of a Calabi-Yau space slowly changes and develops a node at time $t=2$.
Black holes then condense, implementing a smooth transition to a
topologically distinct Calabi-Yau space at time $t=3$.
\end{quote}

The appearance of a massless particle often signals a phase transition. One
may wonder if there is a new phase of string theory characterized by
\begin{equation}
\langle \Phi_{\rm BH}\rangle\not= 0\ ,
\end{equation}
where $\Phi_{\rm BH}$ is the field whose quanta are the degenerating
black holes. This may seem like a difficult question, but in fact the
answer is easily determined using $N=2$ supersymmetry, which fixes the
potential for the field $\Phi_{\rm BH}$. In the simple
conifold singularities described in \cite{mbh}, the answer is no: black
hole condensation is prevented by a quartic potential.

The situation is dramatically different for the more complex conifold
singularities analyzed in collaboration with Brian Greene and Dave
Morrison \cite{gms}.
These singularities correspond to multiple degenerations at
which $P$ 3-cycles degenerate and $P$ black holes come down to zero
mass. The  Wilsonian action at the singularity involves $P$ black hole
fields, $\Phi^A_{\rm BH}$, $A=1, \cdots P$. $N=2$ supersymmetry again
determines the potential $V(\Phi^A_{BH})$. In some cases it is found that
$V$ has flat directions along which black holes can condense!

It might appear that a new branch of the string moduli space has been
discovered. However, there is overwhelming evidence that
$\langle\Phi^A_{\rm BH}\rangle \not= 0$ branches are not new string
vacua. Rather they are a new, dual description of old string vacua. The
spectrum of massless particles in the $\langle\Phi^A_{\rm BH}\rangle
\not= 0$ branches agree in each of the thousands of known examples with
the spectrum of a known Calabi-Yau space. Furthermore, pairs of
Calabi-Yau's which are connected in this manner by black hole
condensation are the same as those pairs previously known from the work
of \cite{cgh} to be connected by a singular conifold transition in which
an $S^3$ is shrunk to zero size and then blown back up as an $S^2$.
Hence black hole condensation in four dimensions corresponds to a
change in the topology of the internal Calabi-Yau, as depicted in figure
3. In general relativity the topology of a manifold cannot change in a
smooth fashion. String theory is an extension of general relativity in which
smooth topology change can occur.

Thousands, and possibly all simply-connected, Calabi-Yau's are connected
by such transitions. In this fashion the plethora of disconnected string
vacua are unified into a smaller number --- possibly one --- of moduli spaces
as illustrated in figure 4.
The long-term aspiration is that, when understood, the dynamics of
supersymmetry breaking will select a preferred point(s) in this space.

In the ``old'', Calabi-Yau, description of the $\langle
\Phi^A_{\rm BH}\rangle \not= 0$ phase, $\Phi^A_{\rm BH}$ is identified as a
field whose quanta are fundamental strings rather than black holes.
Thus under the topology-changing phase transitions,
\begin{eqnarray}
{\rm Black\ Holes} &\to &{\rm Strings}\nonumber\\
{\rm Strings} &\to& {\rm Black\ Holes}\ . \nonumber
\end{eqnarray}
Black holes and strings are dual descriptions of the same entity.
For decades theorists have pursued the idea that elementary particles
are secretly black holes. We have seen that a version of this idea is
realized in string theory. Hence string theory succeeds not only in unifying
all particles and forces with one another, but in unifying them with
black holes as well.
\vskip .13 in
\centerline{\epsfysize=3.00in \epsfbox{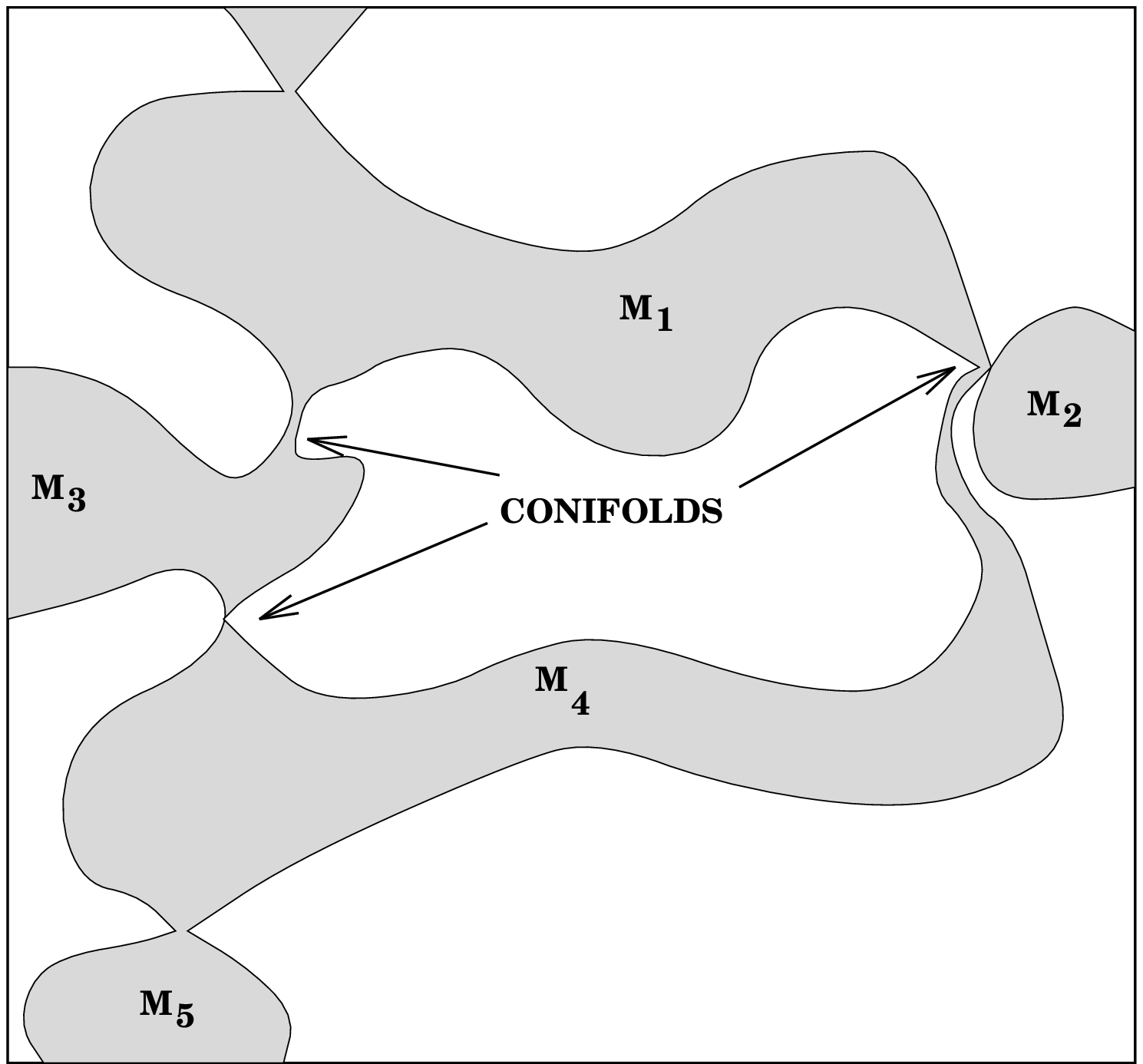}}

\begin{quote}\setlength{\baselineskip}{.1in}{\bf Figure 4}: \sl The
vacuum moduli spaces, $M_1, M_2, \cdots$ of topologically distinct
Calabi-Yau spaces are branches of a larger moduli space connected
via conifold transitions.
\end{quote}

The preceding discussion has close parallels in the beautiful work of Seiberg
and Witten on $N=2$, $d=4$ gauge theories \cite{SW1,SW2}. In the pure
$SU(2)$ gauge theory, \cite{SW1} there is a conifold singularity which
appears at a special point in the moduli space of Higgs $vev$'s.
This theory also contains `t Hooft-Polyakov monopoles which degenerate at
the conifold. The Wilsonian theory including light monopoles is smooth at
the conifold, just like the Wilsonian theory with light black holes
described here.

There are also some apparent differences with the work in \cite{SW1}.
The conifold singularity of \cite{SW1} has an alternate description as a
divergence in the {\it quantum} sum over Yang-Mills instantons, as
opposed to the Calabi-Yau conifolds which have an alternate description
(utilizing mirror symmetry) as a {\it classical} sum over worldsheet
instantons. This distinction evaporates in the context of a dual
description of string theory in which the string is itself a soliton
\cite{dh}. In
such a description the classical worldsheet sum becomes a quantum sum
over spacetime instantons \cite{mbh,klty}.
Explicit examples of this have been understood in the context of dualities
relating type II-heterotic string compactifications \cite{ht}.
Duality promotes the analogy to an identity: the dual transforms of
black holes are monopoles, and worldsheet instantons turn into
Yang-Mills instantons.

Field theory analogs of the conifold transitions in which the topology
of the Calabi-Yau changes also exist \cite{SW2}. For example, in the
$N=2$ $SU(2)$ gauge theory with two flavors of ``quarks'' in an $SU(2)$
doublet, the moduli space has several branches. The first is called
the Coulomb branch, along which $SU(2)$ is  broken to $U(1)$ by an
adjoint Higgs $vev$ and all quarks are massive. At special conifold points on
the Coulomb branch massless charged states appear. These can
condense and form a new branch called the Higgs branch along which the
$U(1)$ is broken. Condensation of these massless charged
states creates a new branch of the gauge theory moduli  space in the
same fashion that black hole condensation creates a new branch of the
string moduli space.

String dualities again promote the analogy to an identity. From the dual,
heterotic perspective, the exotic topology-changing conifold transitions
of the type II theory are
nothing but condensation of various light-charged fields: The moduli
space of $N=2$ heterotic string vacuua contains many special
points where charged perturbative string states
become light and condense, changing both the
massless spectrum and dimension of the moduli space. Hence, a
consistent
picture of heterotic-type II string duality relies crucially
on the existence of
 black hole condensation in type II theories.

Perhaps the most exciting aspect of recent developments is the deep new
puzzles they have raised. I would like to draw attention to one of these
puzzles related to the preceding analogy. Consider a moduli field which
is slowly rolling in a generic fashion and encounters a conifold
singularity. Part of spacetime will spill across the transition, and a
bubble of the new phase will form. Inside the bubble a new
spectrum of massless particles will appear. Our analysis of the
low-energy effective action enables one to obtain the lowest-order
approximation to this process.

However, in a complete theory one should, in principle, be able to
compute arbitrarily high order corrections to the leading approximation.
Clearly, the usual string perturbation rules are useless here because
different conformal field theories are relevant to the regions
inside and outside the
bubble.  We do not have a rule for computing these corrections. It is
furthermore clear that, whatever those rules are, they are quite
different from the usual rules of string theory.

The analogy with the Seiberg-Witten field theory case is again
illuminating.
In that case the low-energy effective theories on the Higgs and Coulomb
branches can be used to give a leading-order description of the
formation of a bubble of the Higgs branch inside the Coulomb branch.
However, a systematic computation of the corrections can only be made
from knowledge of the microscopic $SU(2)$ gauge theory.

In our current understanding of string theory, it is as if we have seen
the last equations in the papers of
Seiberg and Witten which describe the low-energy
effective abelian gauge theories. To fully understand string theory, we
must work backwards from these last equations to the first equations in
which the theory is fundamentally defined as an $SU(2)$ gauge theory.

Clearly this is an enormous task. At the same time, recent developments
have provided us with new tools and concrete questions with  which we
can address
these issues, and progress is being made in leaps and bounds.
It is an exciting time for string theory.
\vskip .13 in
\noindent{\bf ACKNOWLEDGMENTS}

This work was supported in part by a grant from the Department of
Energy, 91ER40618.


\begin{thebibliography}{}

\bibitem{chsw}
P.~Candelas, G.T.~Horowitz, A.~Strominger, and E.~Witten,   Nucl.~Phys.~B,
 258 (1985) 46.

\bibitem{CDGP91} P.~Candelas, X.~dela Ossa, P.~Green, and L.~Parkes,
Nucl.~Phys.~B, 359 (1991) 21.

\bibitem{mbh} A.~Strominger,
Nuc.~Phys.~B,  451 (1995) 96-108;  hep-th/9504090.

\bibitem{yssc} A.~Strominger,
Phys.~Rev.~Lett.,   55 (1985) 2547.

\bibitem{bbs} K.~Becker, M.~Becker and A.~Strominger, hep-th/9507158.

\bibitem{hst} G.\ Horowitz and A.\ Strominger,
Nucl.\ Phys.\ B,  360 (1991) 197.

\bibitem{vfa} C.\ Vafa, hep-th/9505023.

\bibitem{gms}  B.R.~Greene, D.R.~Morrison, and A.~Strominger, Nucl.~Phys.~B,
 451 (1995) 109-120.

\bibitem{mrei87} M.~Reid, Math.~Ann., 278 (1987) 329.

\bibitem{cgh} P.~Candelas, A.~Dale, A.~Lutken, and R.~Schimmrigk, Nuc.~Phys.~B,
298 (1988) 493; P.~Candelas, P.~Green, and T.~Hubsch, Nucl.~Phys.~B,  42
(1990) 246.

\bibitem{SW1} N.~Seiberg and E.~Witten, Nucl.~Phys.~B,  426 (1994) 19.

\bibitem{SW2} N.\ Seiberg and E.\ Witten, Nucl.\ Phys.\ B, 431 (1991)
484.

\bibitem{dh} A.~Dabholkar and J.~Harvey, Phys.~Rev.~Lett., 63 (1989) 478.

\bibitem{klty} A.~Klemm, W.~Lerche, S.~Theisen and S.~Yankielowicz,
Phys.~Lett., B344 (1995) 169, hep-th/9411057.

\bibitem{ht} C.M.~Hull and P.K.~Townsend, Nucl.\ Phys.\ B, 438 (1995)
47, hep-th/9410167;
E.~Witten,  Nucl.\ Phys.\ B, 443 (1995), hep-th/9503124;
S.\ Kachru and C.\ Vafa, hep-th/9505105;
S.~Ferrara, J.A.~Harvey, A.~Strominger, and C.~Vafa,
hep-th/9505152.


\end{thebibliography}
\end{document}